\documentclass[aps,prl,twocolumn,superscriptaddress]{revtex4-1} %for 2 column
\usepackage{graphicx}% Include figure files
\usepackage{dcolumn}% Align table columns on decimal point
\usepackage{bm}% bold math
\usepackage{natbib}
\usepackage{wasysym}
\usepackage{textcomp}
\usepackage{bbm}
\newcommand{\Id}{\mathbbm{1}}
\newcommand{\ket}[1]{{|{#1}\rangle}}
\newcommand{\bra}[1]{{\langle{#1}|}}
\newcommand{\ketbra}[1]{\ket{#1}\bra{#1}}

\begin{document}
%\preprint{}
%
\title{Inductive measurement of optically hyperpolarized phosphorous donor nuclei in an isotopically-enriched silicon-28 crystal}

%\author{}
%\thanks{}
%\affiliation{}
\author{P. Gumann} \email{gumann@physics.harvard.edu}
\affiliation{Institute for Quantum Computing, University of Waterloo, Waterloo, Ontario N2L 3G1, Canada}
\affiliation{Deparment of Physics and Astronomy, University of Waterloo, Waterloo, Ontario N2L 3G1, Canada}
\affiliation{Deparment of Physics, Harvard University, Cambridge, MA 02138, USA}
\author{O. Patange}
\affiliation{Institute for Quantum Computing, University of Waterloo, Waterloo, Ontario N2L 3G1, Canada}
\affiliation{Deparment of Physics and Astronomy, University of Waterloo, Waterloo, Ontario N2L 3G1, Canada}
\affiliation{Waterloo Institute for Nanotechnology, University of Waterloo, Waterloo, Ontario N2L 3G1, Canada}
\author{C. Ramanathan} \email{chandrasekhar.ramanathan@dartmouth.edu}
\affiliation{Department of Physics and Astronomy, Wilder Laboratory, Dartmouth College, Hanover, New Hampshire 03755, USA}
\author{H. Haas}
\affiliation{Institute for Quantum Computing, University of Waterloo,  Waterloo, Ontario N2L 3G1, Canada}
\affiliation{Deparment of Physics and Astronomy, University of Waterloo, Waterloo, Ontario N2L 3G1, Canada}
\author{O. Moussa}
\affiliation{Institute for Quantum Computing, University of Waterloo, Waterloo, Ontario N2L 3G1, Canada}
\affiliation{Deparment of Physics and Astronomy, University of Waterloo, Waterloo, Ontario N2L 3G1, Canada}
\author{M. L. W. Thewalt} 
\affiliation{Department of Physics, Simon Fraser University, Burnaby, British Columbia V5A 1S6, Canada}
\author{H. Riemann} 
\affiliation{Leibniz-Institut fuer Kristallzuechtung, 12489 Berlin, Germany}
\author{N. V. Abrosimov} 
\affiliation{Leibniz-Institut fuer Kristallzuechtung, 12489 Berlin, Germany}
\author{P. Becker} 
\affiliation{PTB Braunschweig, 38116 Braunschweig, Germany}
\author{H.-J. Pohl} 
\affiliation{VITCON Projectconsult GmbH, 07743 Jena, Germany}
\author{K. M. Itoh} 
\affiliation{School of Fundamental Science and Technology, Keio University, Yokohama, 3-14-1 Hiyoshi, 223-8522, Japan}
\author{D. G. Cory}
\affiliation{Institute for Quantum Computing, University of Waterloo, Waterloo, Ontario N2L 3G1, Canada}
\affiliation{Deparment of Physics and Astronomy, University of Waterloo, Waterloo, Ontario N2L 3G1, Canada}
\affiliation{Waterloo Institute for Nanotechnology, University of Waterloo, Waterloo, Ontario N2L 3G1, Canada}
\affiliation{Canadian Institute for Advanced Reserach, Toronto, Ontario M5G 1Z8, Canada}
\affiliation{Department of Chemistry, University of Waterloo, Waterloo, Ontario N2L 3G1, Canada}
\affiliation{Perimeter Institute for Theoretical Physics, Waterloo, Ontario N2L 2Y5, Canada}

\date{\today}

\begin{abstract}

We experimentally demonstrate the first inductive readout of optically hyperpolarized phosphorus-31 donor nuclear spins in an isotopically enriched silicon-28 crystal. The concentration of phosphorus donors in the crystal was 1.5 x 10$^{15}$ cm$^{-3}$, three orders of magnitude lower than has previously been detected via direct inductive detection. The signal-to-noise ratio measured in a single free induction decay from a 1 cm$^3$ sample ($\approx 10^{15}$ spins) was 113.  By transferring the sample to an X-band ESR spectrometer, we were able to obtain a lower bound for the nuclear spin polarization at 1.7~K of $\sim 64$~\%. The $^{31}$P-T$_{2}$ measured with a Hahn echo sequence was 420~ms at 1.7~K, which was extended to 1.2~s with a Carr Purcell cycle. The T$_1$ of the $^{31}$P nuclear spins at 1.7~K is extremely long and could not be determined, as no decay was observed even on a timescale of 4.5 hours. Optical excitation was performed with a 1047~nm laser, which provided above bandgap excitation of the silicon. The build-up of the hyperpolarization at 4.2~K followed a single exponential with a characteristic time of 577~s, while the build-up at 1.7~K showed bi-exponential behavior with characteristic time constants of 578~s and 5670~s.
\end{abstract}

\pacs{}
%\keywords{}
\maketitle

Nuclear spin defects are archetype models of qubits in solid state systems. We expect them to have long coherence times and to be well controlled \cite{Morton-2008,Awschalom-2013}. However, to date they have mainly been studied via their interaction to a neighboring electron spin \cite{Morton-2008,Muhonen-2014,Awschalom-2013,Pla-2013}. Such experiments are indirect probes of the local fields seen by the nuclear spins. Here, we directly observe nuclear spin defects in a dilute sample of silicon and through a combination of FID and echo measurements we characterize the local field and its fluctuations.

%Hybrid electron-nuclear spin systems are of interest in quantum information processing as they offer the possibility of combining fast control and good state preparation and readout of the electron spin, with the long coherence times of the nuclear spin degree of freedom. Nuclear spins offer the promise of useful quantum memories in such spin-based architectures \cite{Morton-2008,Awschalom-2013}. 

The phosphorus donor impurity in silicon is a potentially promising candidate for a hybrid quantum information processor \cite{Kane-1998}. In natural abundance bulk silicon, the 300-600~$\mu$s coherence time of the donor electron spin at low temperatures has been shown to be limited primarily by spectral diffusion due to the $^{29}$Si nuclei (4.7~\% natural abundance) \cite{Tyryshkin-2006}. Similar coherence times have also been measured at the level of individual donors \cite{Pla-2013, Muhonen-2014}. In the bulk, this coherence time has been extended to 0.6~s by isotopically engineering the silicon lattice to reduce the $^{29}$Si nuclear spin concentration and simultaneously reduce the donor concentration to minimize the dipolar coupling between electron spin donors (thus reducing instantaneous diffusion effects) \cite{Tyryshkin-2011}. The $^{31}$P donor nuclear spin has also been shown to have extremely long coherence times (180~s at low temperature and B=845~G) \cite{Steger-2012}, limited primarily by electron spin fluctuations. By ionizing the donors with below-gap narrow-line laser excitation and using dynamical decoupling techniques, the phosphorus nuclear spin coherence times were extended to 39 minutes at room temperature and 3~hours at 4.2~K in a silicon-28 lattice, at $\sim 845$~G \cite{Saeedi-2013}.

It has recently been shown that it is possible to optically hyperpolarize the $^{31}$P donor nuclear spins in silicon at relatively low doping concentrations ($\sim 10^{15}$ cm$^{-3}$) in two different regimes. At high magnetic field ($\sim 8.5$~T), the phosphorus nuclear spins were detected using both electron spin resonance (ESR) and Electrically Detected Magnetic Resonance (EDMR) \cite{Morley-2008,McCamey-2009,McCamey-2012, Lo-2013} under white light illumination. The optical nuclear hyperpolarization of -68~\% built up over a characteristic time of 120~s \cite{McCamey-2009}. Due to the limited penetration of the light into the silicon, the hyperpolarization occurred primarily near the illuminated surface. At low magnetic fields, the nuclear spin polarization (86~\%) was measured using Photoluminesence Excitation (PLE) Spectroscopy with both resonant and above bandgap laser excitation \cite{Yang-2006,Yang-2009,Steger-2012} and showed sub-second optical hyperpolarization timescales.

Here we demonstrate the direct inductive readout of the phosphorus Nuclear Magnetic Resonance (NMR) signal at a  phosphorus donor concentration of $\sim 10^{15}$ cm$^{-3}$ \cite{Dreher-2012}, following hyperpolarization of bulk $^{31}$P nuclei using non-resonant infra-red laser excitation, at high field and low temperature. Previous direct NMR measurements of phosphorus nuclear spins in silicon have only been possible at very high doping concentrations ($\sim 10^{18}$ cm$^{-3}$) \cite{Alloul-1987,Jeong-2009}, about three orders of magnitude higher than the concentrations used in this paper. This inductive readout of the phosphorus donor nuclei allows us to measure nuclear spin properties in the bulk of the sample.

\begin{figure}[!bt]
\centering
\includegraphics[scale=0.3]{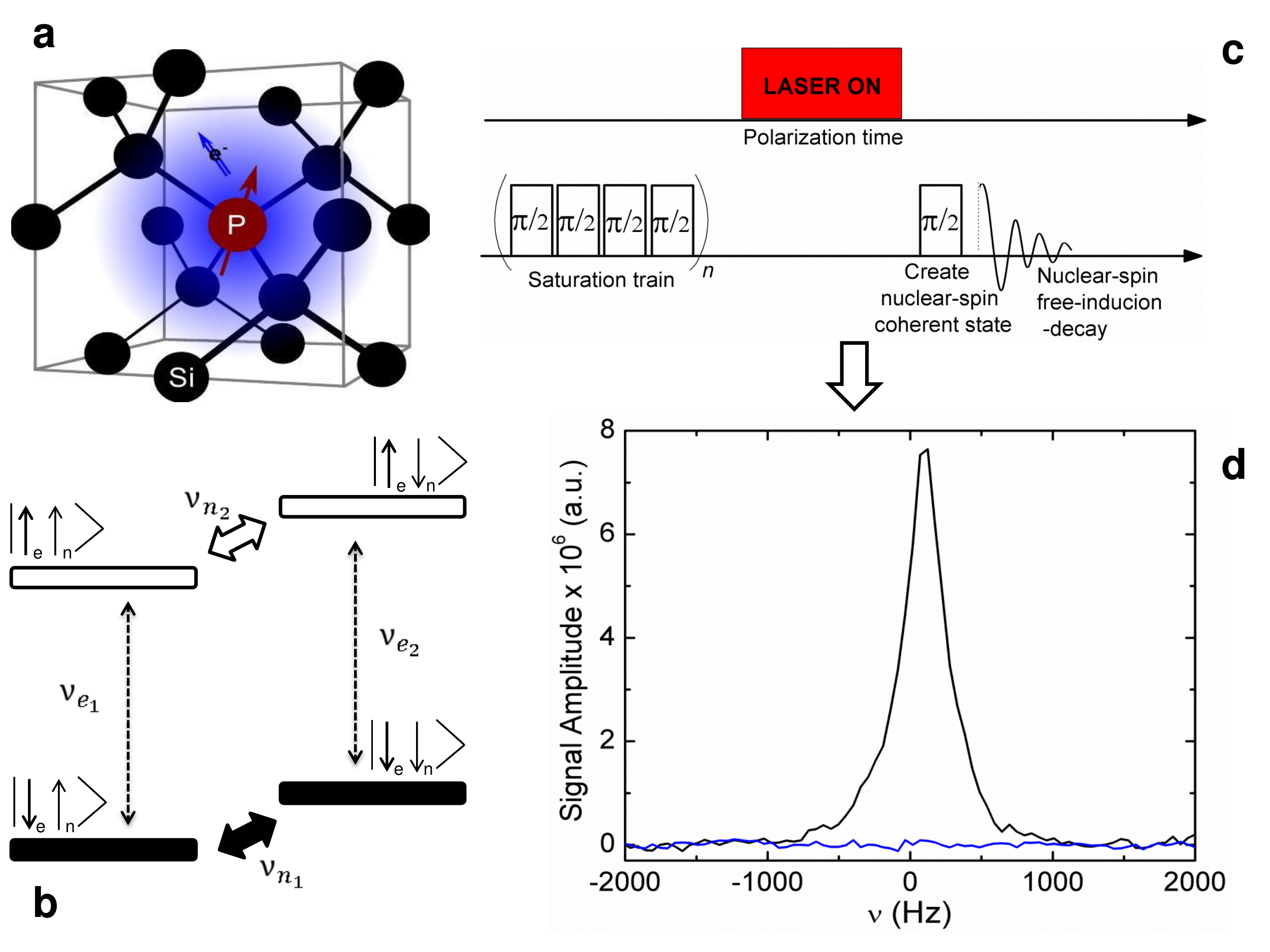}
\caption{Raw spectrum of $^{31}$P-nuclear spins in $^{28}$Si crystal, at 1.7~K in 6.7~T. The data were taken by applying only one $\pi/2$ pulse and recording the FID. SNR is 113. a) Schematic and b) eletronic structure of the donor impurity. c) Pulse sequence used to obtain spectrum in d).}
\label{fig:energy_diagram}
\end{figure}
We used a simple NMR detection setup where a cylindrical $^{28}$Si-enriched crystal \cite{Becker-2010}, with phosphorus concentration of $1.5 \times 10^{15}$ cm$^{-3}$ (boron concentration $\sim 1.0 \times 10^{14}$ cm$^{-3}$, dislocation free crystal) was placed in a rhodium flashed, silver plated copper, \textit{RF}-coil, wired to a low temperature \textit{LC}-circuit. All experiments presented here were performed at temperatures 4.2~K or 1.7~K \textpm 0.3~K and the magnetic field was 6.71~T. The build-up of the high $^{31}$P-spin polarization was accomplished by illuminating the sample with a 100~mW, 1047~nm, above-bandgap laser, with a linearly polarized beam of 8 mm effective size (see Supplementary Information). The (indirect) bandgap in silicon is 1.12~eV which corresponds to an optical wavelength of 1100~nm. The penetration depth for 1047~nm light in silicon at cryogenic temperatures is a few centimeters which allowed the excitation of bulk phosphorus impurities \cite{Macfarlane-1959}. 

The effective Hamiltonian of the phosphorus donor impurity at high magnetic field is:
\begin{equation}
\mathcal{H} = - \gamma_n B_z I_z - \gamma_e B_z S_z + \frac{2\pi}{\hbar} A S_z I_z
\label{eq-Ham}
\end{equation}
where $\gamma_n/2\pi = 17.23$~MHz/T and $\gamma_e/2\pi = -28.024$~GHz/T are the nuclear and electron gyromagnetic ratios, respectively, and $A = 117.54$~MHz is the isotropic hyperfine interaction term. In the high-field limit the eigenstates are almost exactly given by the product states $\mid \uparrow_{e} \uparrow_{n}\rangle$, $\mid \uparrow_{e} \downarrow_{n}\rangle$, $\mid \downarrow_{e} \uparrow_{n}\rangle$, $\mid \downarrow_{e} \downarrow_{n}\rangle$ \cite{Schweiger-2001}, see Figure \ref{fig:energy_diagram}b. At 6.71~T the thermal electron spin polarization is 79~\% at 4.2~K and 99~\% at 1.7~K while the thermal nuclear spin polarization is 0.07~\% at 4.2~K and 0.16~\% at 1.7~K. We probed the nuclear spins in the lower spin electron manifold, transition $\nu_{n_1} = 174.08$~MHz (see Figure 1b).

Figure \ref{fig:energy_diagram}c illustrates the experimental sequence used to measure the build-up of the phosphorus hyperpolarization. Following a saturation train of $\pi/2$ pulses to destroy the remnants of the hyperpolarization from the previous experiment, the nuclear spins are polarized with laser irradiation. The NMR signal was measured using a single $\pi/2$ RF-pulse (duration 8.5~$\mu$s), and the resulting free induction decay was Fourier transformed to produce the NMR spectrum. A typical signal is show in Figure \ref{fig:energy_diagram}d, produced with 200~s laser irradiation. The full line width at half height is $\sim$160~Hz (consistent with T$_{2}^{*}\sim$ 2~ms).

\begin{figure}[!bt]
\centering
\includegraphics[scale=0.4]{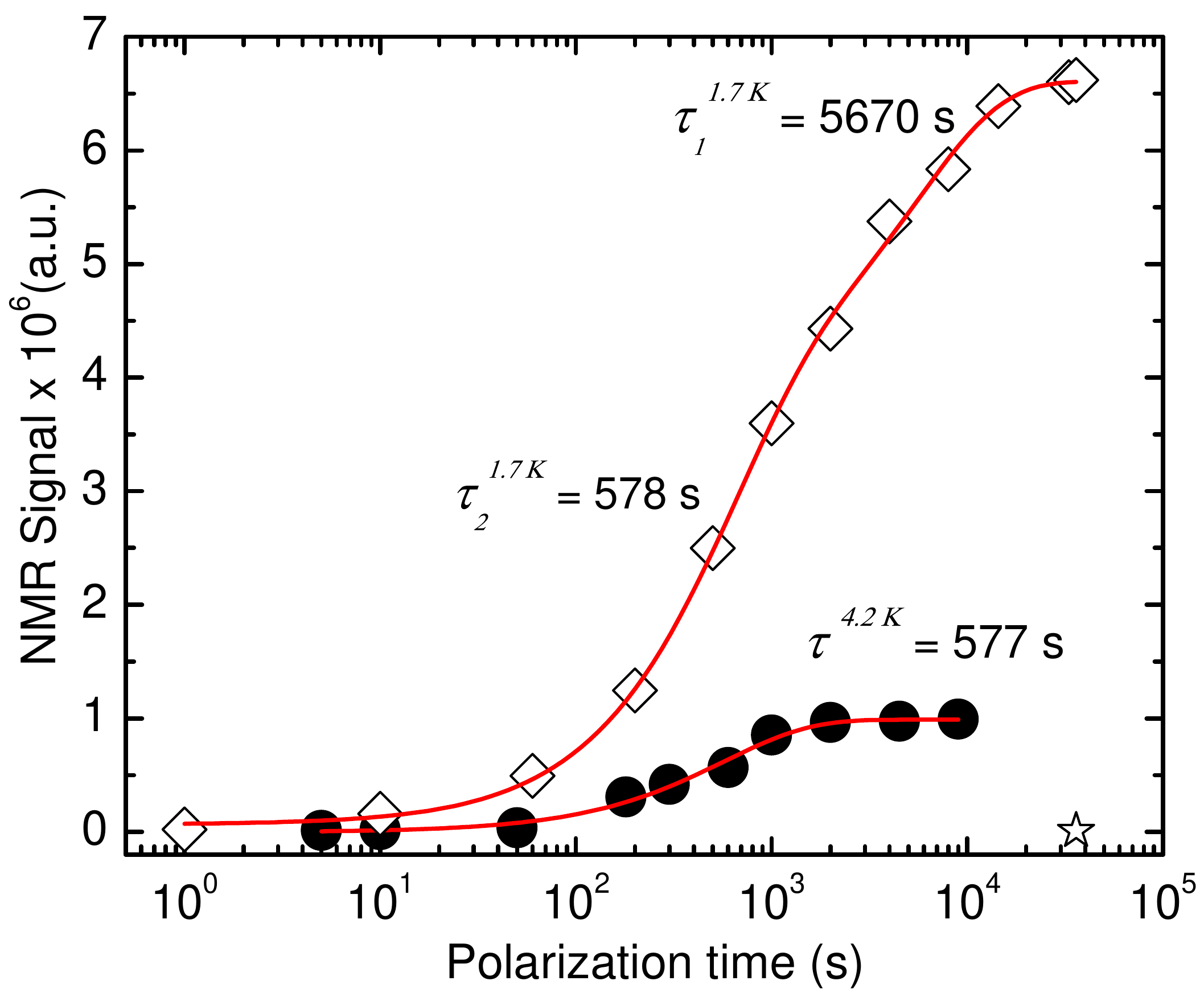}
\caption{Build-up on the nuclear spin polarization by 1047~nm laser irradiation for up to $\sim$10~h, at 1.7~K and 4.2~K temperature respectively. The red lines represent a bi-exponential fit with time constants, $\tau_{1}$ = 5670~s and $\tau_{2}$ = 578~s at 1.7~K, and an exponential fit with $\tau$ = 577~s at 4.2~K. The star represents a thermal polarization measurement (laser off) for $\sim$10~h, at 1.7~K, where no polarization could be observed.} %7->10 hrs for total, 4->7 for T1
\label{fig:build_up}
\end{figure}

The build-up of the hyperpolarization was measured by varying the laser excitation time (or polarization time), from 2~s to 10 hours (Fig. \ref{fig:build_up}). This build-up was measured at both 4.2~K and 1.7~K. The ratio of the steady state signals at these temperatures was measured to be 5.88. We were able to fit the build-up curve at 4.2~K using a single exponential fit with a characteristic time of 577~s. The measured build-up at 1.7~K showed bi-exponential behavior, with characteristic times of 578~s and 5670~s. The relative contributions of the two components were 57.3~\% and 42.7~\% respectively. Comparing the amplitude of the short time constant component at 1.7~K with the signal at 4.2~K, both of which had similar growth times, indicates an enhancement of 3.78. Assuming a simple Boltzmann scaling of the electron spin polarization, lowering the temperature from 4.2~K to 1.7~K should just change the polarization by a factor of 1.25.  

There are at least two contributions to this additional enhancement. First, the efficiency of coupling the laser to the silicon crystal is improved at low temperature as the liquid helium bath enters a superfluid phase below 2.17~K and consequently bubbles in the bath are eliminated. At 4.2~K we infact observe substantial bubbling of the liquid helium at the inner window of the Dewar. These bubbles reduce the effective coupling of the light onto the sample. In addition, the electron spin T$_1$ is longer at low temperature \cite{Tyryshkin-2011}, and the interplay with the optically excited carriers could enhance the polarization \cite{Honig-1970,Thornton-1973}.

%suggested that the conduction electrons and the donors rapidly equilibrate to the same spin polarization following exchange scattering \cite{Honig-1970,Thornton-1973}. 
%If the rate of carrier generation is low, the resulting spin polarization is close to the equilibrium donor polarization, while if the rate of carrier generation is high, the resulting spin polarization tends to zero \cite{Honig-1970}, effectively saturating the electron spin system. 

The detailed physics underlying the optical hyperpolarization process is not well understood. Honig and co-workers have previously shown that the negatively ionized donors produced by spin-trapping of optically-excited conduction band electrons form singlet states at high-field \cite{Thornton-1973}. Similarly, optical experiments have shown the creation of donor-bound excitons at both low \cite{Yang-2009} and high magnetic fields \cite{Sekiguchi-2010}, and the electron-pairs in these donor-bound excitons also form singlets.  When the electron spin polarization (of the donors and free electrons) is high, it is necessary to flip either the donor or the free electron to form the bound singlet.  Sekiguchi {\em et al.\ }have suggested that when spin-orbit interactions are weak, as in silicon, this trapping process is most likely mediated by the hyperfine interaction, resulting in the hyperpolarization of the nuclear spins \cite{Sekiguchi-2010}. Altenatively, the hyperpolarization could be produced by cross-relaxation of the donors, as they are heated up by the optically-excited conduction band electrons \cite{Feher-1959,Pines-1957}.

The long time component of the growth curve observed at 1.7~K was not measured beyond 2.5~h polarization time at 4.2~K (Fig. \ref{fig:build_up}). A similar bi-exponential growth has been observed in a recent microwave-induced DNP experiment on phosphorus donors in natural abundance silicon (doping concentration of $6.5 \times 10^{16}$ cm$^{-3}$) at 4.6~T and temperatures of 200~mK and 1~K \cite{Jarvinen-2014}. They observed a short timescale of 15~s and a longer timescale of 1100~s in their experiment. Though they attribute the presence of the longer timescale to the presence of $^{29}$Si spins around the phosphorus donors, this is unlikely to be the case here, as a similar bi-exponential behavior is observed in our isotopically-enriched silicon-28 crystal.  

We were unable to measure the signal from the phosphorus nuclei in the absence of hyperpolarization, making it difficult to directly quantify either the sign or the magnitude of the nuclear spin polarization. In order to estimate the phosphorus nuclear spin polarization, we moved the sample to an X-band CW ESR spectrometer following optical excitation at 4.2~K for three hours at 6.71~T. The resulting ESR spectrum, measured at 4.2~K, is shown in Fig. 2, Supplementary Material. The magnitude of the measured phosphorus polarization, calculated from the difference in the integrated intensities of the two ESR lines, is -11~\%, which is the lower bound for the induced hyperpolarization at 4.2~K, as some of the polarization will have decayed as the sample was removed from the 6.71~T field and warmed up, before being cooled back down in the ESR cryostat. This indicates a lower bound of $\sim 64$~\% ($11\times 5.88$) for the polarization at 1.7~K. The negative sign of the hyperpolarization, indicated by the higher intensity of the high-field line compared to the low-field line, is in agreement with prior high-field EDMR results \cite{Morley-2008,McCamey2-2012}.

\begin{figure}
\centering
\includegraphics[scale=0.4]{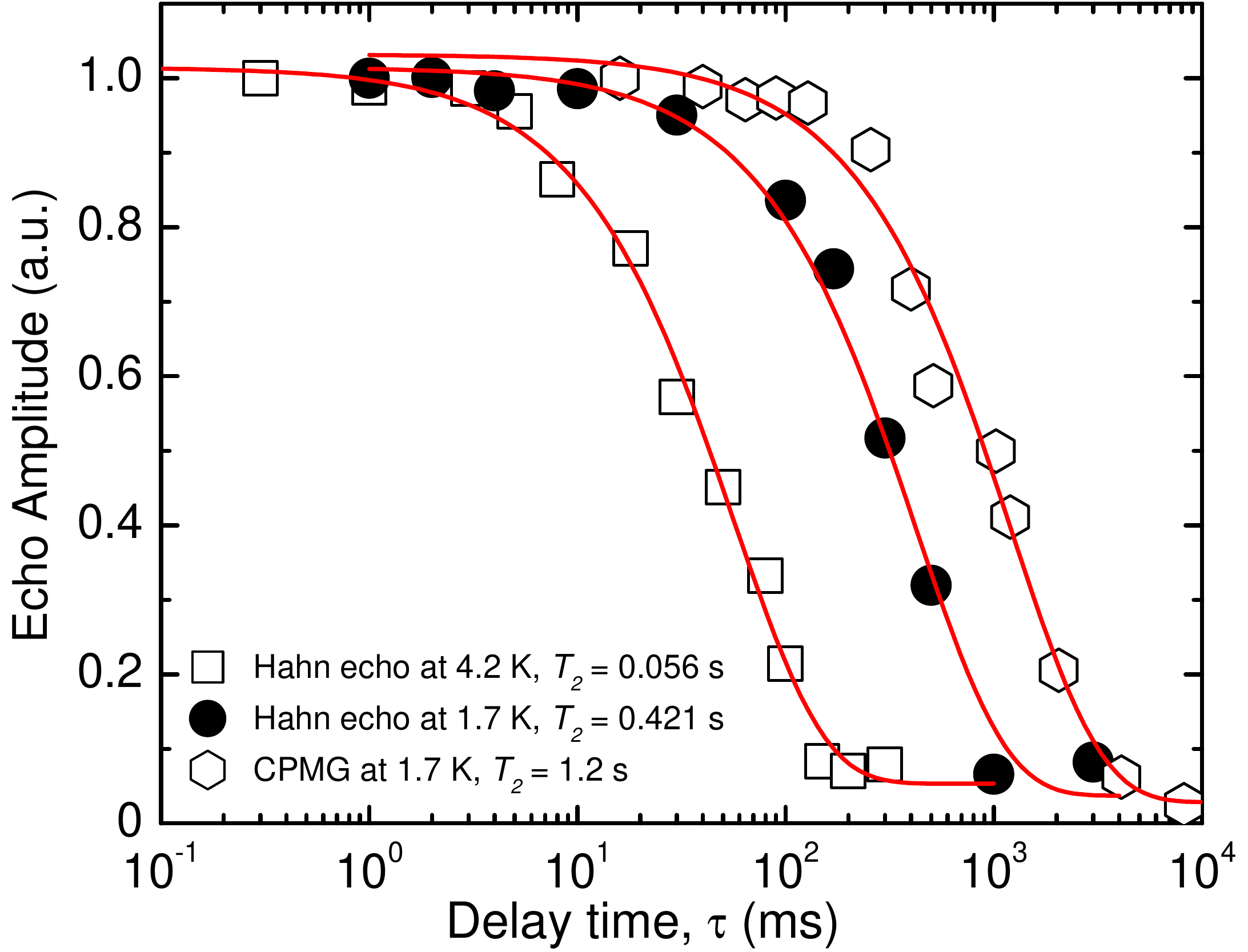}
\caption{Nuclear spin coherence time, T$_{2}$, measured with the Hahn echo at 4.2~K temperature, open squares, Hahn echo at 1.7~K, full circles, and CPMG pulse sequence at 1.7~K, open diamonds. All data were measured in 6.7~T field, with 200~s of optical polarization provided by a 1047~nm, 100~mW, above-gap laser.}
\label{fig:T2}
\end{figure}

We performed spin-echo experiments to measure the coherence time of the $^{31}$P nuclear spins. Following 200~s of laser irradiation, a Hahn-echo sequence ($\pi/2-\tau-\pi-\tau-$acquire) was used to measure the nuclear spin coherence time (Fig. \ref{fig:T2}). By recording the echo signal while varying the delay time ($\tau$), we measured the signal decay at both 4.2~K and 1.7~K as shown in Figure \ref{fig:T2}. We fit the data with a single exponential decay, and measured nuclear spin T$_{2}$ values of 56~ms and 421~ms at 4.2~K and 1.7~K respectively.

As the magnetic field is increased, it is observed that the electron spin T$_1$ at low temperature, and high field gets significantly shorter since T$_1^{-1} \propto B^{4}$ as the result of a direct single-phonon relaxation process \cite{Honig-1958,Honig-1960, Roth-1960, Hasegawa-1960}. The hyperfine interaction is field independent so the main factor limiting the nuclear T$_2$ is the electron T$_1$ carrying the $^{31}$P-spin to the electron spin $\mid \uparrow_{e} \rangle$ manifold \cite{note2} (see Supplementary Materials for details). In the presence of light, the T$_1$ is further shortened by up to two orders of magnitude due to trapping and re-emission, with T$_1$ on the order of 2~ms in the presence of light and almost 20~ms in the dark at 8.56~T \cite{Morley-2008,McCamey2-2012}.

Here the electron spin undergoing T$_1$ relaxation induces an effective T$_2$ process on the nuclear spin with time constant T$_2^{en}$(see Supplementary Information). If $AT^{e}_1 \gg$ 1 then
\[
T_2^{en} = \frac{T^{e}_1}{p_\uparrow} ,
\]
where $p_\uparrow$ is the probability for the electron to be in the excited state. The high temperature limit of this model has been applied to explain the nuclear T$_2$ \cite{Morton-2008}. If we assume that the experimentally observed nuclear $T_2(T)$ combines two independent effects $1/ T_2(T) = 1/T_2' + p_\uparrow(T) / T^{e}_1(T)$, where $T_2'$ is temperature independent, then we obtain $1 / T_2' \le 1 / T_2(\text{1.7~K})$.
This in turn puts an upper bound on the electron relaxation time $T^{e}_1(\text{4.2~K}) \le  p_\uparrow(\text{4.2~K}) \frac{T_2(\text{1.7~K})~ T_2(\text{4.2~K})}{T_2(\text{1.7~K})~ - ~T_2(\text{4.2~K})}$ or $T^{e}_1(\text{4.2~K}) \le 6.7~ \text{ms}$, where we have assumed that $p_\uparrow$ is given by the equilibrium thermal probability.  This value is shorter than the T$_1^e=20$~ms measured in the dark at 8.56~T \cite{McCamey2-2012}.

In order to minimize the effect of environmental fluctuations we applied a CPMG refocusing pulse sequence to extend the nuclear spin coherence time. In the CPMG sequence the single $\pi$ pulse of the Hahn echo is substituted with a series of $\pi$ pulses that are 90\textdegree out of phase with respect to each other, with a $\tau$ spacing of 2~ms. The resulting echo decay is presented in Figure \ref{fig:T2}, with a single exponential fit to the data returning T$_{2}$ = 1.2~s \textpm 0.1~s, a factor of almost 3 improvement in nuclear spin coherence time. This is similar to the value of 1.75~s measured previously using ENDOR at 5.5~K \cite{Morton-2008}. This CPMG sequence  will refocus interactions between the phosphorus nucleus and other spins (or fields) that are fluctuating on a time scale longer than a few hundred Hertz. The sequence will thus refocus fluctuations due to distant donor electrons, silicon nuclei (the silicon-phosphorus nuclear dipolar coupling is very small and does not play an important role here \cite{note}) and static field inhomogeneities. The phosphorus nuclear dipolar coupling is not refocused, but is only about 1.5~mHz for our donor concentration, and the dominant contribution from the  electron T$^{e}_1$ induced nuclear T$_2$ is also not refocused by the CPMG sequence.

\begin{figure}
\centering
\includegraphics[scale=0.4]{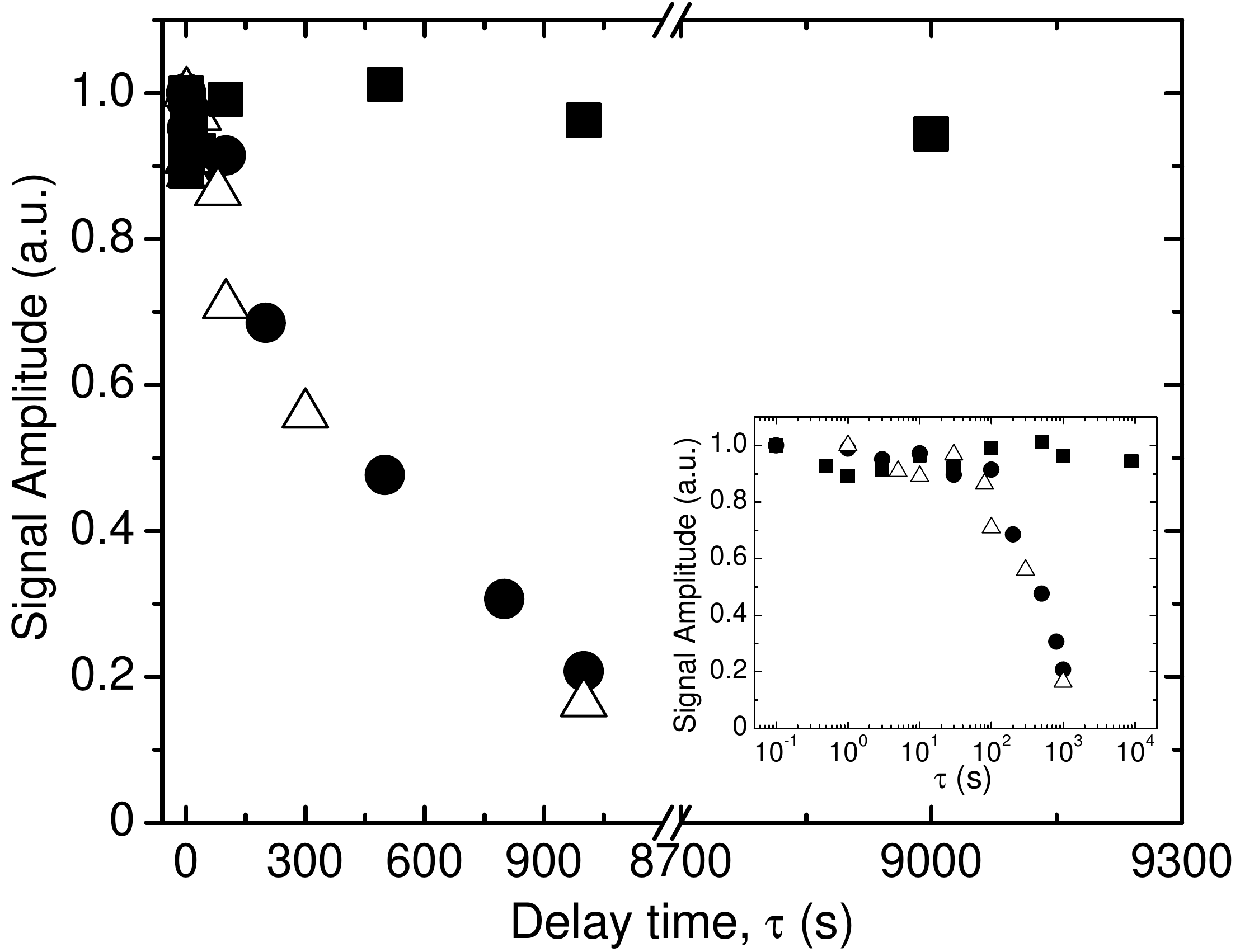}
\caption{Nuclear spin relaxation time, T$_{1}$, measured with a simple $\pi/2$ read-out pulse at 4.2~K (open triangles) and 1.7~K (black squares), respectively, and with an inversion recovery pulse sequence (closed circles) at 4.2~K.}
\label{fig:T1}
\end{figure}

Lastly, we confirmed the long T$_{1}$ relaxation times, at 4.2~K and 1.7~K temperatures. Figure \ref{fig:T1} shows T$_{1}$ data for two experiments, a 200~s laser polarization pulse, followed by: in the first case a delay time $\tau$ and a $\pi/2$-read-out pulse; in the second case $\pi$-$\tau$-$\pi/2$ pulse sequence. The only difference between the two runs is the initial nuclear state. If most of the population is localized in $\mid \downarrow_{e} \uparrow_{n}\rangle$ state, applying a $\pi$-pulse before the read-out pulse will move it to the $\mid \downarrow_{e} \downarrow_{n}\rangle$ state (Fig. \ref{fig:energy_diagram}b). The T$_{1}$ relaxation should not depend on the initial state, which is confirmed in Fig. \ref{fig:T1}. In addition we observe that the spin-lattice relaxation time not only increases at lower temperatures but also exceeds the measuring times of our setup, no visible decay was observed after waiting for delay time $\tau=$4.5~h (Fig. \ref{fig:T1}). 

In conclusion, the results presented here show the first single FID measurement of the local magnetic fields seen by $^{31}$P nuclear spins in a dilute crystal of  $^{28}$Si. The negative $^{31}$P polarization is $> $11~\% at 4.2~K, and $>$64 \% at 1.7~K and 6.71~T. It was accomplished by directly illuminating the sample with an above gap 1047~nm laser for over 5~h at 1.7~K and 2.7~h at 4.2~K. We were able to extend the T$_{2}$ relaxation time to 1.2~s at 1.7~K, and confirm an extremely long T$_1$ of the $^{31}$P nuclear spins at 1.7~K which could not be determined within the timescale of this experiment. 
%The build-up of the hyperpolarization at 4.2~K demonstrated a single exponential recovery with a characteristic time of 577~s, while the build-up at 1.7~K showed bi-exponential behavior with characteristic time constants of 578~s and 5670~s.

This work was supported by the Natural Sciences and Engineering Research Council of Canada (NSERC), Canadian Excellence Research Chairs (CERC) Program and the Canadian Institute for Advanced Research (CIFAR) and Industry Canada. C.R. acknowledges support of a Junior Faculty Fellowship from Dartmouth College.

The $^{28}$Si-enriched sample used in this study was prepared from Avo28 material produced by the International Avogadro Coordination (IAC) Project (2004-2011) in cooperation among the BIPM, the IN-RIM (Italy), the IRMM (EU), the NMIA (Australia), the NMIJ (Japan), the NPL (UK), and the PTB (Germany).

\section{Supplementary Material}

\subsection{Nuclear spin $T_2$ due to electron spin undergoing $T_1$ relaxation}
We consider an electron-nuclear spin system coupled via $S_z I_z$ interaction Hamiltonian. If the electron spin in the system is undergoing a $T_1$ relaxation process the nuclear spin will experience an effective Hamiltonian that switches randomly between $S_z$ and $-S_z$ as the electron spin fluctuates between its excited state $\ket{\uparrow}$ and ground state $\ket{\downarrow}$. This randomly fluctuating effective field will induce a dephasing $T_2$ process on the nuclear spin. Here we provide a calculation for that nuclear $T_2$ mechanism taking an open quantum systems approach \cite{BreuerPetruccione} modelling the electron relaxation process with the help of Lindblad operators \cite{Lindblad, BrasilLindblad, PearleLindblad}, a similar, albeit less general, calculation has been given by Morton {\sl et al.} \cite{Morton-2008}.

%We say that we have a system of nuclear and electron spins, there is no inherent relaxation mechanism for the nuclear spin (this may or may not be the case and it would not ), the electron spin is undergoing an asymmetric (this is to account for near polarised equilibrium density matrices) spin flip (T1) relaxation meaning that in any narrow time window $\Delta t$ the electron has probability of flipping to up state and probability of flipping to down state. 

For the following derivation we assume that we have an electron-nuclear spin system fully described by a density matrix $\rho_{en}$ which evolves under the Hamiltonian defined in Equation (1) of the main text. We recast the Hamiltonian to a more convenient form
\begin{equation}
\label{eq:Hamiltonian1}
\mathcal{H} = \frac{\hbar \omega_e}{2} ~ \sigma_z \otimes \Id - \frac{\hbar \omega_n}{2} ~ \Id \otimes \sigma_z + \frac{\hbar \omega_{en}}{4} ~ \sigma_z \otimes \sigma_z ,
\end{equation}
where $\sigma_z = \ketbra{\uparrow}-\ketbra{\downarrow}$ is the Pauli $z$ operator and $\Id = \ketbra{\uparrow}+\ketbra{\downarrow}$ is the identity operator, while $\omega_e = -\gamma_e B_z$, $\omega_n = \gamma_n B_z$, $\omega_{en} = 2 \pi A$. We assume that the electron $T_1$ is the only relaxation mechanism for the coupled spin system, hence the nuclear $T_1 = \infty$ and the electron $T_2$ is purely a result of its $T_1$ process, yet it should be noted that adding an additional $T_2$ mechanism for the electron spin would not alter the conclusions of this calculation as we will later assume that at no point during the experiment will the electron spin have coherences. The electron $T_1$ relaxation is modelled using Lindblad equation which dictates the dissipative time evolution of the electron density matrix $\rho_e$ according to
\begin{eqnarray}
%-\frac{i}{\hbar} [\frac{\hbar \omega_e}{2} ~ \sigma_z, \rho_e] \\
\nonumber \frac{\partial}{\partial t} \rho_{e}^{diss} &=-\frac{1}{2 T_1}~ (1-p_\uparrow)~ \left[ \sigma_+ \sigma_- \rho_e + \rho_e \sigma_+ \sigma_- -2 \sigma_- \rho_e \sigma_+ \right] \\
&- \frac{1}{2 T_1}~ p_\uparrow~ \left[ \sigma_- \sigma_+ \rho_e + \rho_e \sigma_- \sigma_+ -2 \sigma_+ \rho_e \sigma_- \right],
\label{eq:Lindblad1}
\end{eqnarray}
where $\sigma_+ = \ket{\uparrow} \bra{\downarrow}$, $\sigma_- = \ket{\downarrow} \bra{\uparrow}$ and $\rho_e = \textup{Tr}_n [\rho_{en}]$ is the reduced density matrix for the electron spin found by tracing out the nuclear spin. $\frac{\partial}{\partial t} \rho_{e}^{diss}$ denotes the time derivative of $\rho_{e}$ due to dissipative effects only, the full dynamics of $\rho_{e}$ would be found by combining $\frac{\partial}{\partial t} \rho_{e}^{diss}$ with Hamiltonian dynamics under $\mathcal{H}$ in Equation (\ref{eq:Hamiltonian1}) and tracing out the nuclear spin. Equation (\ref{eq:Lindblad1}) describes an asymmetric relaxation process with characteristic time scale $T_1$ driving the density matrix $\rho_e$ towards the equilibrium $\rho_{eq} = p_\uparrow~\ketbra{\uparrow}+(1-p_\uparrow)~\ketbra{\downarrow}$. It is easy to show that for $\rho_{eq}$ the derivative in Equation (\ref{eq:Lindblad1}) vanishes and it could be thought of as the thermal equilibrium density matrix for the electron spin, $p_\uparrow$ being the probability for the spin to be in the excited state $\ket{\uparrow}$. Equation (\ref{eq:Lindblad1}) is a convex combination of two continuously acting amplitude damping processes \cite{MikeIke}, the first line weighted by probability $(1-p_\uparrow)$ drives the electron spin state towards $\ket{\downarrow}$ while the second line weighted by probability $p_\uparrow$ drives the electron spin state towards $\ket{\uparrow}$.

Electron-nuclear density matrix $\rho_{en}$ evolves under the Hamiltonian in Equation (\ref{eq:Hamiltonian1}) and Lindblad operators in Equation (\ref{eq:Lindblad1}), the latter are taken to act only on the electron state i.e. they act as an identity on the nuclear part of the density matrix (reduced density matrix for the nuclear spin is given by partial trace $\textup{Tr}_e [\rho_{en}]$). This is a valid assumption as long as the electron $T_1$ relaxation results from electron spin couplings to its environment that act as an identity on the nuclear spin. We deduce that the electron nuclear density matrix $\rho_{en}$ evolves under
\begin{eqnarray}
\nonumber \frac{\partial}{\partial t} \rho_{en} &=-\frac{i}{\hbar} [\mathcal{H}, \rho_{en}] + \frac{1}{T_1}~ (1-p_\uparrow) (\sigma_- \otimes \Id) \rho_{en} (\sigma_+ \otimes \Id) \\
\nonumber &-\frac{1}{2 T_1}~ (1-p_\uparrow)~ \left[ (\sigma_+ \sigma_- \otimes \Id) \rho_{en} + \rho_{en} (\sigma_+ \sigma_- \otimes \Id) \right] \\
\nonumber &-\frac{1}{2 T_1}~ p_\uparrow~ \left[ (\sigma_- \sigma_+ \otimes \Id) \rho_{en} + \rho_{en} (\sigma_- \sigma_+ \otimes \Id) \right] \\
\label{eq:LiouvilleEq1}
&+ \frac{1}{T_1}~ p_\uparrow (\sigma_+ \otimes \Id) \rho_{en} (\sigma_- \otimes \Id) .
\end{eqnarray}
The experiment described in the main text was performed in the electron ground state $\ket{\downarrow}$ manifold i.e. all excitation pulses and measurements were carried out at angular frequency $\omega_n + \frac{\omega_{en}}{2}$, accordingly we move into the rotating frame of the nuclear Hamiltonian $\mathcal{H}_n = -\left(\frac{\hbar \omega_n}{2} + \frac{\hbar \omega_{en}}{4} \right) ~ \Id \otimes \sigma_z$ and look at the evolution of the transformed density matrix $\tilde{\rho}_{en}(t) = e^{\frac{i \mathcal{H}_n t}{\hbar}} ~ \rho_{en} ~ e^{- \frac{i \mathcal{H}_n t}{\hbar}}$ which determines our experimental observables. Differentiating $\tilde{\rho}_{en}$ and using the definitions in Equations (\ref{eq:Hamiltonian1}) and (\ref{eq:LiouvilleEq1}), while noting that all operators commute with each other, it is easy to show that
\begin{eqnarray}
\nonumber \frac{\partial}{\partial t} \tilde{\rho}_{en} &=-\frac{i}{\hbar} [\tilde{\mathcal{H}}, \tilde{\rho}_{en}] + \frac{1}{T_1}~ (1-p_\uparrow) (\sigma_- \otimes \Id) \tilde{\rho}_{en} (\sigma_+ \otimes \Id) \\
\nonumber &-\frac{1}{2 T_1}~ (1-p_\uparrow)~ \left[ (\sigma_+ \sigma_- \otimes \Id) \tilde{\rho}_{en} + \rho_{en} (\sigma_+ \sigma_- \otimes \Id) \right] \\
\nonumber &-\frac{1}{2 T_1}~ p_\uparrow~ \left[ (\sigma_- \sigma_+ \otimes \Id)\tilde{\rho}_{en} + \tilde{\rho}_{en} (\sigma_- \sigma_+ \otimes \Id) \right] \\
\label{eq:LiouvilleEq2}
&+ \frac{1}{T_1}~ p_\uparrow (\sigma_+ \otimes \Id) \tilde{\rho}_{en} (\sigma_- \otimes \Id) ,
\end{eqnarray}
where $\tilde{\mathcal{H}}$ is the effective Hamiltonian in the rotating frame,
\begin{eqnarray}
\nonumber \tilde{\mathcal{H}} &= \frac{\hbar \omega_e}{2} ~ \sigma_z \otimes \Id + \frac{\hbar \omega_{en}}{4} ~ \left( \Id \otimes \sigma_z + \sigma_z \otimes \sigma_z \right) \\
\nonumber &= \frac{\hbar \omega_e}{2} ~ \sigma_z \otimes \Id + \frac{\hbar \omega_{en}}{2} ~ \ketbra{\uparrow} \otimes \sigma_z .
\end{eqnarray}
It is evident that the effective nuclear spin Hamiltonian is an operator conditional on the electron spin state: if the electron is in its ground state $\ket{\downarrow}$ the effective nuclear Hamiltonian reduces to $0$, whereas if the electron is in the excited state $\ket{\uparrow}$ the nuclear spin experiences the Hamiltonian $\frac{\hbar \omega_{en}}{2} \sigma_z$. At times when the electron spin becomes excited under its $T_1$ process the nuclear spin will accumulate random phases evolving under the Hamiltonian $\frac{\hbar \omega_{en}}{2} \sigma_z$ leading to a $T_2$ process on the nuclear spin.

We notice that evolution under Equation (\ref{eq:LiouvilleEq2}) cannot create coherences between electron spin states $\ket{\downarrow}$ and $\ket{\uparrow}$, meaning that Equation (\ref{eq:LiouvilleEq2}) takes electron-nuclear density matrices of form
\begin{eqnarray}
\nonumber \tilde{\rho}_{en}(t) &= [1-p(t)]~ \ketbra{\downarrow} \otimes \tilde{\rho}_{n \downarrow}(t) \\
\label{eq:reducedrho}
&+p(t)~ \ketbra{\uparrow} \otimes \tilde{\rho}_{n \uparrow}(t)
\end{eqnarray}
only to density matrices of the same form. $p(t)$ in Equation (\ref{eq:reducedrho}) is a time dependent scalar value obeying $0 \le p(t) \le 1$, while $\tilde{\rho}_{n \downarrow}(t)$ and $\tilde{\rho}_{n \uparrow}(t)$ represent time dependent 'conditional' nuclear density matrices for respective electron spin states $\ket{\downarrow}$ and $\ket{\uparrow}$. Assuming that we begin the experiment with no electron coherences we can reduce our problem of keeping track of all 16 real parameters determining $\tilde{\rho}_{en}(t)$ to keeping track of 8 real parameters. We define an 8 dimensional vector $\vec{r}(t)$ which combines two 4 dimensional vectors like $\vec{r}(t) = (\vec{r}_\downarrow (t),\vec{r}_\uparrow (t))$, with $\vec{r}_{\downarrow} = (r_{I_{\downarrow}}, r_{x_{\downarrow}}, r_{y_{\downarrow}}, r_{z_{\downarrow}})$ and $\vec{r}_{\uparrow} = (r_{I_{\uparrow}}, r_{x_{\uparrow}}, r_{y_{\uparrow}}, r_{z_{\uparrow}})$, and rewrite the electron-nuclear density matrix $\tilde{\rho}_{en} (t)$ of interest as
\begin{eqnarray}
\nonumber \tilde{\rho}_{en} (t) &= \frac{1}{\sqrt{2}} \ketbra{\downarrow} \otimes \vec{r}_\downarrow (t). ( \Id, \sigma_x, \sigma_y, \sigma_z) \\
\label{eq:reducedrhovec}
&+ \frac{1}{\sqrt{2}} \ketbra{\uparrow} \otimes \vec{r}_\uparrow (t). ( \Id, \sigma_x, \sigma_y, \sigma_z) .
\end{eqnarray}
$( \Id, \sigma_x, \sigma_y, \sigma_z)$ in Equation (\ref{eq:reducedrhovec}) is a vector of operators as $\sigma_x = \ket{\uparrow} \bra{\downarrow} + \ket{\downarrow} \bra{\uparrow}$ and $\sigma_y = -i \ket{\uparrow} \bra{\downarrow} + i \ket{\downarrow} \bra{\uparrow}$ are the Pauli operators. We note that Equation (\ref{eq:reducedrhovec}) is merely a convenient reparametrization of $\tilde{\rho}_{en}(t)$ in  Equation (\ref{eq:reducedrho}) using an orthonormal set of operators, the idea of the reparametrization is similar to that of the damping basis introduced in \cite{BriegelEnglert}, $\vec{r}_\downarrow (t)$ and $\vec{r}_\uparrow (t)$ are the Bloch vectors for the 'conditional' nuclear density matrices $\tilde{\rho}_{n \downarrow}(t)$ and $\tilde{\rho}_{n \uparrow}(t)$ rescaled by probabilities $[1-p(t)]$ and $p(t)$.

The time evolution of $\tilde{\rho}_{en}(t)$ in Equations (\ref{eq:reducedrho}) and (\ref{eq:reducedrhovec}) is fully determined by the time evolution of $\vec{r}(t)$. Differentiating both sides of Equation (\ref{eq:reducedrhovec}) and employing the orthonormality and Hermicity of the operators we deduce that
\[
\dot{\vec{r}} = \frac{1}{\sqrt{2}} \left(
\begin{array}{c}
 \textup{Tr} \left[ \left( \ketbra{\downarrow} \otimes \Id \right)  \frac{\partial}{\partial t} \tilde{\rho}_{en} \right] \\
 \textup{Tr} \left[ \left( \ketbra{\downarrow} \otimes \sigma_x \right)  \frac{\partial}{\partial t} \tilde{\rho}_{en} \right] \\
 \textup{Tr} \left[ \left( \ketbra{\downarrow} \otimes \sigma_y \right)  \frac{\partial}{\partial t} \tilde{\rho}_{en} \right] \\
 \textup{Tr} \left[ \left( \ketbra{\downarrow} \otimes \sigma_z \right)  \frac{\partial}{\partial t} \tilde{\rho}_{en} \right] \\
 \textup{Tr} \left[ \left( \ketbra{\uparrow} \otimes \Id \right)  \frac{\partial}{\partial t} \tilde{\rho}_{en} \right] \\
 \textup{Tr} \left[ \left( \ketbra{\uparrow} \otimes \sigma_x \right)  \frac{\partial}{\partial t} \tilde{\rho}_{en} \right] \\
 \textup{Tr} \left[ \left( \ketbra{\uparrow} \otimes \sigma_y \right)  \frac{\partial}{\partial t} \tilde{\rho}_{en} \right] \\
 \textup{Tr} \left[ \left( \ketbra{\uparrow} \otimes \sigma_z \right)  \frac{\partial}{\partial t} \tilde{\rho}_{en} \right] \\
\end{array} 
\right).
\]
Substituting the definition of $\tilde{\rho}_{en}(t)$ in Equation (\ref{eq:reducedrhovec}) into the right hand side of Equation (\ref{eq:LiouvilleEq2}) and using the resulting expression for $\frac{\partial}{\partial t} \tilde{\rho}_{en}$ we can evaluate the vector elements of $\dot{\vec{r}}$ which are best expressed as a matrix equation
\[
\dot{\vec{r}} = \left(
\begin{array}{cccccccc}
 \frac{-p_\uparrow}{T_1} & 0 & 0 & 0 & \frac{1-p_\uparrow}{T_1} & 0 & 0 & 0 \\
 0 & \frac{-p_\uparrow}{T_1} & 0 & 0 & 0 & \frac{1-p_\uparrow}{T_1} & 0 & 0 \\
 0 & 0 & \frac{-p_\uparrow}{T_1} & 0 & 0 & 0 & \frac{1-p_\uparrow}{T_1} & 0 \\
 0 & 0 & 0 & \frac{-p_\uparrow}{T_1} & 0 & 0 & 0 & \frac{1-p_\uparrow}{T_1} \\
 \frac{p_\uparrow}{T_1} & 0 & 0 & 0 & \frac{p_\uparrow-1}{T_1} & 0 & 0 & 0 \\
 0 & \frac{p_\uparrow}{T_1} & 0 & 0 & 0 & \frac{p_\uparrow-1}{T_1} & -\omega_{en} & 0 \\
 0 & 0 & \frac{p_\uparrow}{T_1} & 0 & 0 & \omega_{en} & \frac{p_\uparrow-1}{T_1} & 0 \\
 0 & 0 & 0 & \frac{p_\uparrow}{T_1} & 0 & 0 & 0 & \frac{p_\uparrow-1}{T_1} \\
\end{array}
\right).\vec{r} .
\]
The linear differential equation for $\vec{r}(t)$ can be solved in the regime of $\omega_{en} T_1 \gg 1$ which is the regime for our experimental parameters, in such case we can approximate 
$\vec{r}(t) \approx \exp(M t).\vec{r}(0)$, where 
\[
M = \\
\left(
\begin{array}{cccccccc}
 \frac{-p_\uparrow}{T_1} & 0 & 0 & 0 & \frac{1-p_\uparrow}{T_1} & 0 & 0 & 0 \\
 0 & \frac{-p_\uparrow}{T_1} & 0 & 0 & 0 & 0 & 0 & 0 \\
 0 & 0 & \frac{-p_\uparrow}{T_1} & 0 & 0 & 0 & 0 & 0 \\
 0 & 0 & 0 & \frac{-p_\uparrow}{T_1} & 0 & 0 & 0 & \frac{1-p_\uparrow}{T_1} \\
 \frac{p_\uparrow}{T_1} & 0 & 0 & 0 & \frac{p_\uparrow-1}{T_1} & 0 & 0 & 0 \\
 0 & 0 & 0 & 0 & 0 & \frac{p_\uparrow-1}{T_1} & -\omega_{en} & 0 \\
 0 & 0 & 0 & 0 & 0 & \omega_{en} & \frac{p_\uparrow-1}{T_1} & 0 \\
 0 & 0 & 0 & \frac{p_\uparrow}{T_1} & 0 & 0 & 0 & \frac{p_\uparrow-1}{T_1} \\
\end{array}
\right) .
\]
Having set the four matrix entries of $M$ to $0$ amounts to a secular approximation of ignoring all entries that do not commute with the entries of $\omega_{en}$. Matrix $M$ is easily exponentiated by noticing that reshuffling its rows and columns turns $M$ into a block diagonal matrix with each block being just a $2 \times 2$ matrix.

We take the initial electron-nuclear density matrix to be given by $\tilde{\rho}_{en}(0) = \ketbra{\downarrow} \otimes \left( \frac{1}{2} \Id + r_x \sigma_x + r_y \sigma_y + r_z \sigma_z \right)$ corresponding to $\vec{r}(0)=\sqrt{2} \left(\frac{1}{2},r_x, r_y, r_z,0,0,0,0 \right)$, where $r_x$, $r_y$, $r_z$ are the nuclear Bloch vector components after having applied the excitation pulses. The excitation pulses are assumed to be applied on time scales shorter than $T_1$ and $\tilde{\rho}_{en}(0)$ is taken to have no $\ketbra{\uparrow} \otimes \vec{r}_\uparrow.( \Id, \sigma_x, \sigma_y, \sigma_z)$ part since the pulses leave the electron excited state manifold invariant so it yields no observable signal. Evaluating $\exp(M t)$ as described above enables us to find
\[
\vec{r}(t) = \sqrt{2} \left(
\begin{array}{c}
 \frac{1}{2} \left[1 - p_\uparrow \left( 1 - e^{-\frac{t}{T_1}} \right) \right] \\
 e^{-\frac{p_\uparrow t}{T_1}} r_x \\
 e^{-\frac{p_\uparrow t}{T_1}} r_y \\
 \left[1 - p_\uparrow \left( 1 - e^{-\frac{t}{T_1}} \right) \right] r_z \\
 \frac{p_\uparrow}{2} \left(1 - e^{-\frac{t}{T_1}} \right) \\
 0 \\
 0 \\
 p_\uparrow \left(1 - e^{-\frac{t}{T_1}} \right) r_z \\
\end{array} 
\right)
\]
corresponding to
\begin{eqnarray}
\label{eq-finalstate}
\nonumber &\tilde{\rho}_{en}(t) = \ketbra{\downarrow} \otimes \left(e^{-\frac{p_\uparrow t}{T_1}} r_x \sigma_x + e^{-\frac{p_\uparrow t}{T_1}} r_y \sigma_y  \right) \\
\nonumber &+ \left[1 - p_\uparrow \left( 1 - e^{-\frac{t}{T_1}} \right) \right] \ketbra{\downarrow} \otimes \left( \frac{1}{2} \Id + r_z \sigma_z \right) \\
\nonumber &+ p_\uparrow \left( 1 - e^{-\frac{t}{T_1}} \right) \ketbra{\uparrow} \otimes \left( \frac{1}{2} \Id + r_z \sigma_z \right) .
\end{eqnarray}
Tracing out the electron state in the equation above reveals that the reduced density matrix for the nuclear spin at time $t$ is $\tilde{\rho}_{n}(t) = \frac{1}{2} \Id + e^{-\frac{p_\uparrow t}{T_1}} r_x \sigma_x + e^{-\frac{p_\uparrow t}{T_1}} r_y \sigma_y + r_z \sigma_z$ confirming that the electron spin $T_1$ process yields an effective $T_2$ process for the nuclear spin with time constant $T_2 = \frac{T_1}{p_\uparrow}$. Finally, if we assume the equilibrium electron density matrix $\rho_{eq}$ to be determined by Boltzmann distribution at temperature $T$ then $\rho_{eq} = \frac{1}{\textup{Tr} \left[ \exp \left(- \frac{\omega_e \hbar}{k_B T} \frac{\sigma_z}{2} \right) \right]} \exp \left(- \frac{\omega_e \hbar}{k_B T} \frac{\sigma_z}{2} \right)$ and $p_\uparrow = 1/\left(1 + e^{\frac{\omega_e \hbar}{k_B T}}\right)$.

\subsection{T$_{2}$ relaxation with the laser ON}

\begin{figure}
\centering
\includegraphics[scale=0.4]{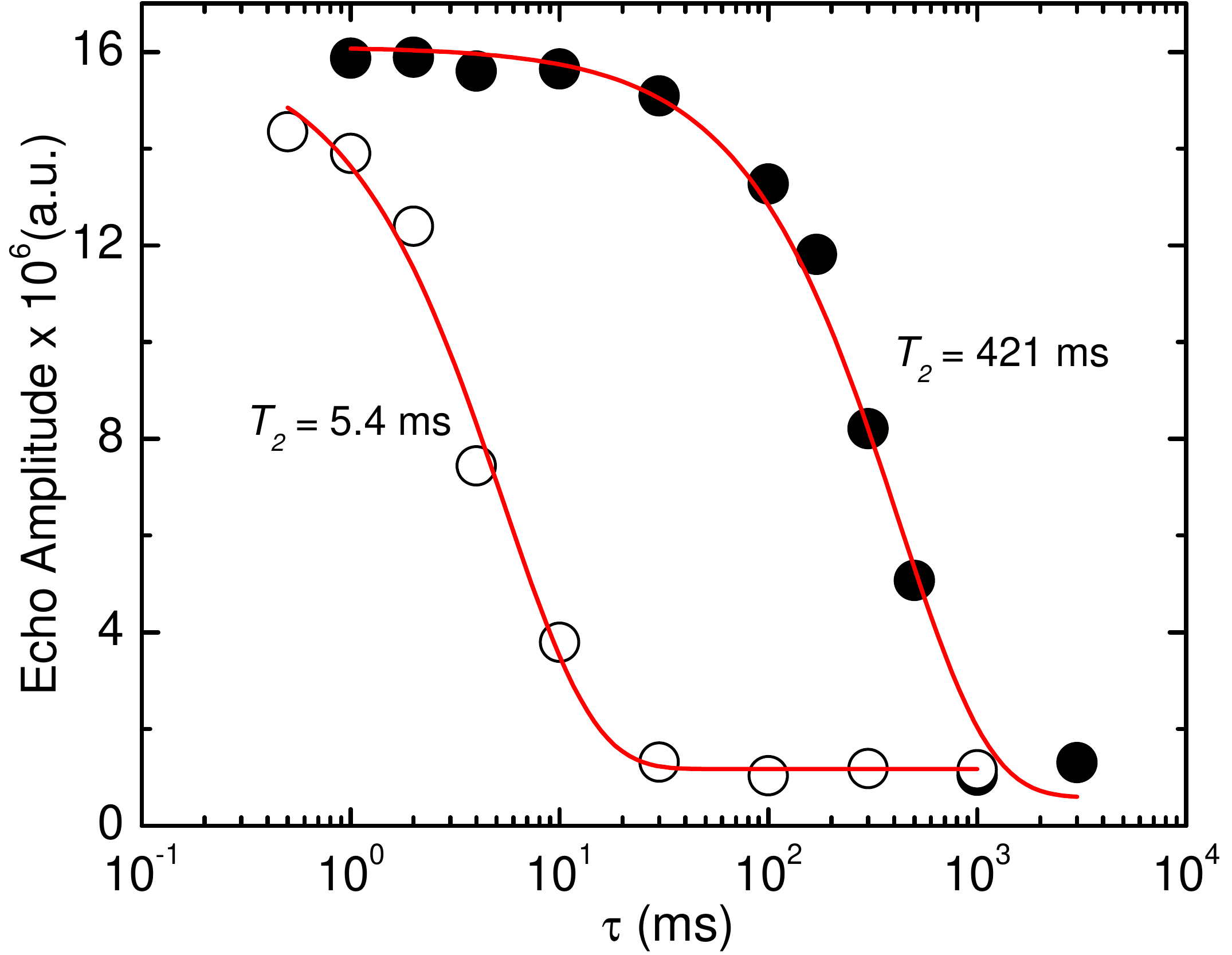}
\caption{Nuclear spin coherence time, T$_{2}$, measured with the Hahn echo at 1.7~K temperature in 6.7~T field. Closed symbols represent a situation where the laser is OFF during the acquisition. Open symbols show data taken using the same pulse sequence but keeping the laser ON during the acquisition.}
\label{fig:T2_with_laserON}
\end{figure} 

In addition to the above mentioned results, we performed an experiment where the laser light was kept on during the Hahn echo pulse sequence. As a result a reduction in T$_{2}$ time, down to 5.4~ms could be observed, Fig. \ref{fig:T2_with_laserON}, which is 78 times shorter than without the light. This indicates that electron relaxation is one of the main sources limiting the nuclear spin T$_{2}$. The measured T$_2$ time scale is consistent with the spin-dependent trapping and re-emission at donor sites that occurs in the presence of the optical excitation, which was observed to result in an electron T$_1$ time of about 2.4~ms [9] under white light irradiation. 

\subsection{Optical Excitation Setup}

The optical excitation setup consists of a light source which is a continuous-wave 1047~nm laser MIL-III-1047 (Opto Engine LLC). The maximum power is 100~mW with up to 100~kHz modulation capabilities. The laser light is linearly polarized with the beam size of 1.6~mm at the laser output, which is then increased by a two-lens telescope to $\sim$ 8~mm. The 45 degrees mirror mounted directly underneath the 6.7~T magnet and centered with its bore directed the laser beam straight onto the sample, through a set of quartz optical windows mounted in a liquid helium Janis cryostat.

\subsection{NMR Experimental Setup}

The sample was placed in an NMR coil mounted in a simple optical cavity to maximize the amount of light irradiation. The home-built, low temperature NMR probe was located inside a liquid helium cryostat (with pumping capabilities) which sits in the bore of a superconducting magnet ($B_0$=6.71~T, with corresponding $^{31}$P resonance frequencies of  $\nu_{n_1} = 174$~MHz and  $\nu_{n_2} = 56$~MHz). The $^{31}$P NMR signals are recorded with a commercial Bruker Avance-300 spectrometer.

\subsection{X-band CW ESR Setup}

Electron spin resonance was undertaken in a commercial X-band (10~GHz) Bruker EMX (Premium X) ESR spectromenter with a Oxford Instruments ESR900 flow cryostat. 

\begin{figure}[!bt]
\centering
\includegraphics[scale=0.4]{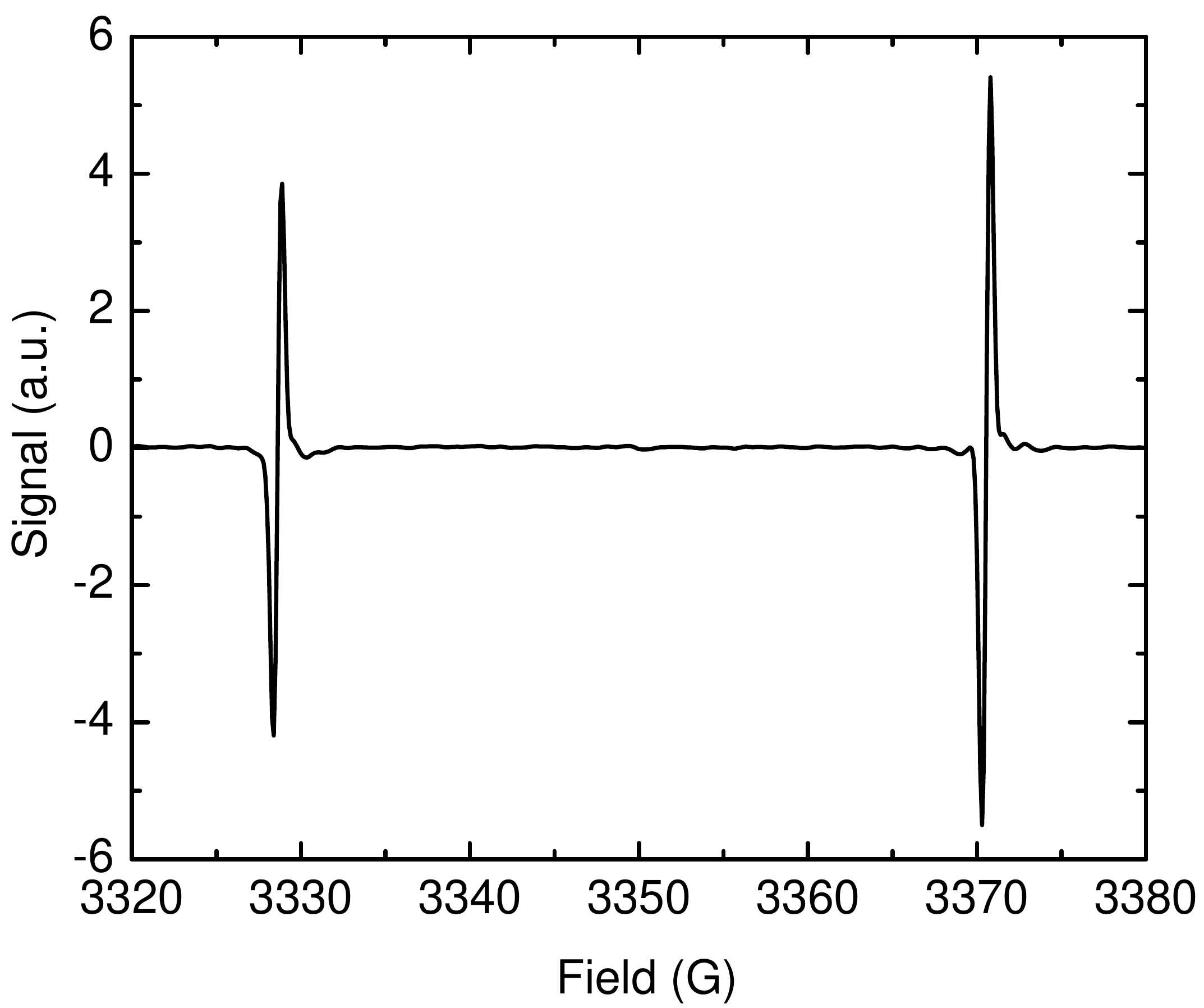}
\caption{Baseline corrected CW ESR spectrum measured using a Bruker X-band, 9.37 GHz, CW ESR spectrometer, following polarization in 6.71~T field at 4.2~K for 3~hours.}
\label{fig:CW-ESR}
\end{figure}

The resulting ESR spectrum, measured at 4.2~K, is shown in Fig. \ref{fig:CW-ESR}. The magnitude of the measured phosphorus polarization, calculated from the difference in the integrated intensities of the two ESR lines, is -11~\%, which is the lower bound for the induced hyperpolarization at 4.2~K, as some of the polarization will have decayed as the sample was removed from the 6.71~T field and warmed up, before being cooled back down in the ESR cryostat. This indicates a lower bound of $\sim 64$~\% ($11\times 5.88$) for the polarization at 1.7~K.

\subsection{Sample description}

The sample was cut from crystal 28Si-10Pr10.6.1PeFZ3, and the growth direction is 100. The phosphorus content is $1.5 \times 10^{15}$ cm$^{-3}$. It was grown from charge 10 material, so enriched to 99.995~\% $^{28}$Si and containing around 46~ppm $^{29}$Si. It is dislocation free. The phosphorus was introduced by adding a small amount of phosphine gas to the argon during the final float-zone single crystal growth run. The sample also contains around $1 \times 10^{14}$ cm$^{-3}$ of boron.


\begin{thebibliography}{99}
\bibitem{note2}
We are in the processes of extending the $^{31}$P T$_2$ by moving the polarized system to higher temperature where the T$_1$ of the electron becomes much less important to the defect T$_2$.
\end{thebibliography}
\end{document}